\documentclass[trackchanges]{aastex701}
\usepackage{amsmath,amssymb,amsfonts}%
\usepackage{mathrsfs}%
\usepackage{xcolor}%

\usepackage{nicefrac,xfrac}

\begin{document}

\title{Structure-wide dark matter density depletion induced by local degeneracies}

\author{Yifei Yang}
\affiliation{South-Western Institute for Astronomy Research, Key Laboratory of Survey Science of Yunnan Province, Yunnan University, Kunming, Yunnan 650500, People's Republic of China}
\email{yangyifei1@stu.ynu.edu.cn}

\author[0000-0003-2240-7031]{Weikang Lin}
\affiliation{South-Western Institute for Astronomy Research, Key Laboratory of Survey Science of Yunnan Province, Yunnan University, Kunming, Yunnan 650500, People's Republic of China}
\email[show]{weikanglin@ynu.edu.cn}  

\correspondingauthor{Weikang Lin}

\begin{abstract}

The longstanding cusp–core problem—the discrepancy between the steep central density cusps predicted by cold dark matter (DM) simulations and certain shallow cores observed in dwarf galaxies, in particular the associated diversity of inner profiles—remains hotly debated despite decades of study. Building on a new interpretation of fermionic isothermal halos, we identify a physical mechanism—degeneracy-induced depletion—in which degenerate inner cores of fermionic DM suppress the surrounding density over large scales. This effect persists even in dense baryonic environments. Within the framework of hierarchical structure formation, degeneracies developed in the smallest constituent subhalos induce low-density regions that collectively configure into a King-type core of the host DM halo, with a core density–radius relation consistent with observations. This scenario accounts for the diversity of DM inner profiles through variation in the average degeneracy of constituent subhalos, and suggests a connection between this diversity and the halo formation history. Thus, the cusp–core problem may be reconciled within the standard “cold” DM paradigm without invoking strong baryonic feedback, instead pointing to the fermionic nature of DM. 

\end{abstract}


\section{Introduction}\label{sec:intro}
Accumulated observations suggest that the DM halos can be cored for dwarf galaxies and low-surface-brightness (LSB) galaxies \citep{Moore1994,Flores:1994gz,deBlok:2008wp,Oh:2015xoa,Rodrigues:2017vto}. This differs from the results of DM-only N-body simulations that predict cuspy profiles at the center \cite[hereafter NFW]{Navarro:1995iw}. Such a discrepancy is called the ``cusp–core'' problem \citep{deBlok:2009sp,Salucci:2018hqu,Boldrini:2021aqk,Sales:2022ich}. Although some simulations argue that baryonic feedback can produce cored DM profiles \citep{DiCintio:2013qxa,Freundlich-etal:2020}, the pronounced diversity of inner DM profiles for dwarf with similar baryon mass remains a major challenge for such feedback-based explanations \citep{Oman:2015xda,Relatores:2019-II,Sales:2022ich}.

New DM physics are motivated to resolve the cusp–core problem \citep{DelPopolo:2021bom}, such as fuzzy DM \citep{Hu-Barkana-Gruzinov-PhysRevLett2000} and DM self-interactions \citep{Spergel:1999mh}. But those proposals remain controversial. For example, it is argued that light boson DM cannot correctly account for the core properties \citep{Deng-etal-2018-PhysRevD,Chavanis:2018pkx,Burkert:2020laq,Banares-Hernandez:2023axy} and the required self-interaction strength is constrained by other considerations \citep{Randall:2008ppe,Hayashi:2020syu}. Nonetheless, the cusp–core problem is thought to disfavor collisionless massive DM candidates, if complicated baryon effects are not involved. This work shall show the opposite.

One fundamental limitation of current N-body simulations is their incapability to capture the quantum statistical effects. This is rooted in the fact that particles in simulations are distinguishable, not to mention the unrealistically huge test-particle mass. 
Recently, a number of works have investigated the role played by fermionic quantum statistics on DM halos at galaxy scales \citep{Destri:2012yn,deVega:2013jfy,2013-Destri-DeVega-Sanchez-DMHalo,Domcke:2014kla,RAR-2015,2015-Chavanis-Lemou-Mehats-II,Randall:2016bqw,DiPaolo:2017geq}. 
In a regime where DM is degenerate at the center, the DM density profiles are cored. Most efforts were devoted to this core, which is caused by the Fermi degenerate pressure preventing the system from gravitational collapse \citep{PhysRev.55.374}.
Unfortunately, to account for the observed core size ranging from hundreds of pc to a few tens of kpc \citep{deBlok:2008wp,Rodrigues:2017vto}, the required DM mass is about one to two orders of magnitude smaller than the mass lower bound obtained from the cosmic large-scale structure  \citep{Bar:2021jff,Carena:2021bqm,Zhang:2024vkx}.

There is more to the degenerate regime: the DM profile exhibits a low-density but extended plateau surrounding the high-density core  \citep{RAR-2015,2015-Chavanis-Lemou-Mehats-II}. Owing to its dominant mass relative to the compact inner core, this plateau is more relevant for interpreting rotation curve observations in dwarf-sized galaxies  \citep{Siutsou:2015}, and it has been observed by \citet{Chavanis:2021jds} that it may account for the observed constant surface density  \citep{2009MNRAS.397.1169D,Burkert:2015-scaling,Salucci:2018hqu}. We refer to the degenerate core as the ``inner core'' while the extended plateau as the ``outer core.''

We examine the properties of the outer core in greater detail and find remarkable agreement between model predictions and observations in the core density–radius relation, particularly when applying an empirically motivated scaling between galactic baryon mass and DM velocity dispersion (Sec.~\ref{sec:outer-core}). However, rather than directly supporting the outer core as the origin of the observed DM cores, this consistency suggests that the observed DM cores resemble classical King-type cores and may arise as an indirect consequence of a depletion mechanism. This motivates us to focus not on the outer core itself, but on the sharp density drop induced by the degenerate inner core. The significance of this feature in shaping the inner profiles and diversity of DM halos, as well as its potential connection to cosmic structure formation, has not been recognized in the literature. We show that the presence of a local degenerate core necessarily produces such a sharp drop in density at its edge (Sec.~\ref{sec:the-DID-mechanism}), and that this behavior is robust even in dense baryonic environments (Sec.~\ref{sec:stability-against-baryon}). We propose that this feature—a mechanism we term Degeneracy-Induced Depletion (DID)—creates low-density regions on galactic scales, which subsequently assemble through hierarchical structure formation and configure under self-gravity into the classical King-type cores responsible for the observed DM halo cores (Sec.~\ref{sec:DID-hierarchical-formation}). This scenario paves a potential new pathway to explain the diversity of DM inner halo profiles through differences in their structure formation histories. (also see Sec.~\ref{sec:DID-hierarchical-formation}).

\section{Fermionic isothermal DM halo}\label{sec:FIP}
We highlight here the key steps in obtaining the fermionic isothermal profiles (FIPs) (see more details in \citet{Bilic-etal:2002,Destri:2012yn,RAR-2015,2015-Chavanis-Lemou-Mehats-II,Randall:2016bqw,Chavanis:2021jds}), adopting the natural units where $c=\hbar=k_{\textsc{b}}=1$. We consider a spherical halo of fermionic DM in dynamical equilibrium,
\begin{equation}\label{eq:Hydro-equil}
    \frac{dP_{\rm d}}{dr}=-\rho_{\rm d} \frac{d\phi}{dr}\,,
\end{equation}
and DM follows a nonrelativistic Fermi-Dirac distribution presumably due to violent relaxation \citep{10.1093/mnras/136.1.101,Chavanis:1998},
\begin{equation}\label{eq:HiDM-distribution}
    f(p) = \frac{g}{2\pi^2}\frac{p^2}{\exp\big[(\frac{p^2}{2m_{\rm d}}-\mu_{\rm eff})/T_{\rm d}\big]+1}\,,
\end{equation}
where $g$ is taken to be $2$ but can be generalized. The effective chemical potential is
\begin{equation}\label{eq:TF-approximation}
    \mu_{\rm eff}=\mu-m_{\rm d}\phi\,,
\end{equation}
where $\phi$ is the gravitational potential and $\mu$ is the real chemical potential. Assuming there is no net particle flow at any radius, we take $\mu$ to be spatially independent. This treatment corresponds to the Thomas-Fermi approach applied in \citet{Bilic-etal:2002,Destri:2012yn}. This assumption is expected to break down at large radii $r$, but it should be a good approximation in the region of interest, as isothermal dark matter profiles are often adopted to describe the inner regions of halos (e.g., see \citet{2008gady.book}). Nonetheless, this assumption is further supported by the overall agreement between the predicted core density–radius relation and observations, as we will show.

The mass density and pressure are calculated in a standard statistical mechanics way and it can be derived that \citep{Zhang:2024vkx},
\begin{align}
    \rho_{\rm d}&=\sqrt{2}Q_*\left(\frac{T_{\rm d}}{m_{\rm d}}\right)^{\frac{3}{2}}J_{\nicefrac{3}{2}}^{\textsc{f}}(\Delta)\,, \label{eq:rho-SM}\\
    P_{\rm d}&= \frac{2^{3/2}Q_*}{3}\left(\frac{T_{\rm d}}{m_{\rm d}}\right)^{\frac{5}{2}}J_{\nicefrac{5}{2}}^{\textsc{f}}(\Delta)\,,\label{eq:pressure-SM}
\end{align}
where $\Delta\equiv\mu_{\rm eff}/T_{\rm d}$ is the degree of degeneracy, $Q_*=\frac{gm_{\rm d}^4}{2\pi^2}$ and $J_s^{\textsc{f}}(\Delta)=\int_0^\infty\frac{q^{s-1}{\rm d}q}{\exp(q-\Delta)+1}$. A useful relation is $\frac{dJ_s^{\textsc{f}}(\Delta)}{d\Delta}=(s-1)J_{s-1}^{\textsc{f}}(\Delta)$ for $s\geq1$. Eq.~\eqref{eq:HiDM-distribution} reduces to the classical Maxwell-Boltzmann distribution when $\Delta\ll-1$ and describes a degenerate fermionic system when $\Delta>0$.

The last equation is Poisson's equation for the gravitational potential,
\begin{equation}\label{eq:Poisson-eq}
    \frac{d\phi}{dr}=\frac{GM_{\rm d}(r)}{r^2}\,.
\end{equation}
We have ignored the general relativistic correction (see \citet{Bilic-etal:2002,RAR-2015,Alberti:2019xaj}). 

With $\mu$ being a constant, substituting Eqs.~\eqref{eq:TF-approximation}, \eqref{eq:rho-SM} and \eqref{eq:pressure-SM} into Eq.~\eqref{eq:Hydro-equil}, it follows that the resulting halo is isothermal, i.e., $T_{\rm d}$ is spatially independent. So, the solutions are fermionic isothermal spheres. Then, with $T_{\rm d}$ being a constant, substituting Eq.~\eqref{eq:Poisson-eq} into \,\eqref{eq:Hydro-equil} and taking the derivative of both sides yields the following second-order differential equation for $\Delta$ that fully describes the DM density profile (see details in \ref{sec:derivation}),
\begin{equation}\label{eq:Delta-differential}
    \Delta''+\frac{2}{z}\Delta'=-J_{\nicefrac{3}{2}}^{\textsc{f}}(\Delta)\,,
\end{equation}
where $'$ denotes the derivative with respect to $z\equiv r/r_*$. The characteristic radius $r_*$ is,
\begin{align}\label{eq:rs}
    r_*&=\frac{2\pi}{g^{1/2}}\left(\frac{m_{\rm d}}{2T_{\rm d}}\right)^{1/4}\left(\frac{m_{\rm p}}{m_{\rm d}}\right)^2\ell_{\rm p}\,,\nonumber\\
    &=\left(\frac{2}{g}\right)^{\frac{1}{2}}\left(\frac{30\,\text{km/s}}{\sigma_v}\right)^{\frac{1}{2}}\left(\frac{5\,\text{keV}}{m_{\rm d}}\right)^2\times0.23\,{\rm pc}\,,
\end{align}
where $\sigma_v=\sqrt{T_{\rm d}/m_{\rm d}}$ is the (1-D) DM velocity dispersion, $m_{\rm p}=\sqrt{\frac{1}{8\pi G}}=2.4\times10^{24}$\,keV the reduced Planck mass and $\ell_{\rm p}=\frac{1}{m_{\rm p}}=8.2\times10^{-35}$\,m. Note that $r_*$ is macroscopic for a wide range of massive DM candidates, although still small compared to kpc scales.

\begin{figure}
    \centering
    \includegraphics[width=0.65\linewidth]{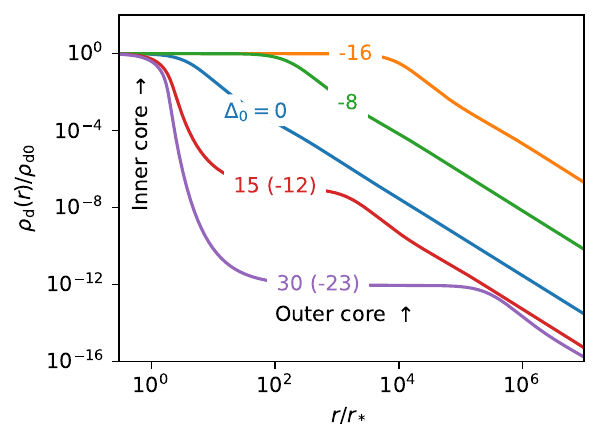}
    \caption{Halo density profiles normalized by the central density. Curves are labeled by the central degree of degeneracy $\Delta_0$. The double-core feature can be clearly seen for profiles with $\Delta_0>0$. When an outer core appears, the value of $\Delta_{\rm oc}$ is shown in parentheses. 
    }
    \label{fig:mass-profiles}
\end{figure}

The solutions to Eq.~\eqref{eq:Delta-differential} are characterized by the central degeneracy, denoted as $\Delta_0$. Resultant DM profiles are shown in Figure~\ref{fig:mass-profiles} presented as the normalized density versus $r/r_*$ with their values of $\Delta_0$ labeled on the curves. The profiles obtained are consistent with previous studies \citep{Destri:2012yn,deVega:2013jfy,2013-Destri-DeVega-Sanchez-DMHalo,RAR-2015,2015-Chavanis-Lemou-Mehats-II,Randall:2016bqw,Chavanis:2021jds}. Two classes of profile are readily identified, one with $\Delta_0<0$ and the other with $\Delta_0>0$. Profiles with $\Delta_0<0$ are nothing but the classical cored isothermal sphere obtained by imposing a finite central density \citep{S.W.cosmo.}. Profiles with $\Delta_0>0$ have a high-density but compact inner core followed by a low-density but extended outer core. It has been shown that the outer core can explain the individual rotation curve of LSB galaxies \citep{Siutsou:2015}. As more cored profiles of dwarf-size galaxies are determined, a universal scaling relation between the core density and radius is reported and supported by increasing observations \citep{2009MNRAS.397.1169D,Burkert:2015-scaling,Salucci:2018hqu}.  We first further examine the properties of the outer core, in particular its relation between density and radius, then we focus on the sharp density drop feature that leads to such an outer-core region.

\section{The outer core and its core density-radius relation}\label{sec:outer-core}
The density profile of the outer core and beyond is in a region where DM has a classical Maxwell-Boltzmann distribution and $J_{\nicefrac{3}{2}}^{\textsc{f}}(\Delta)\simeq \frac{\sqrt{\pi}}{2}e^\Delta$. It implies that the outer core radius should behave like a King's radius such that it is correlated with the outer core density. This can be clearly seen with the solution to Eq.~\eqref{eq:Delta-differential} in the outer core region, which reads
\begin{equation}\label{eq:solution-outer-core}
    \Delta\simeq\Delta_{\rm oc}+\frac{B}{z}-\frac{\sqrt{\pi}}{12}e^{\Delta_{\rm oc}}z^2\,,
\end{equation}
where the constants $\Delta_{\rm oc}$ and $B$ are determined by patching the outer solution with the inner solution. The second term decreases with $z$ quickly and can be ignored. 
The outer core radius $r_{\rm oc}$ can be estimated by the location at which the third term becomes unity, that is, $z_{\rm oc}\propto\exp(-\Delta_{\rm oc}/2)$ and thus $\frac{r_{\rm oc}}{r_*}\propto \frac{1}{\sqrt{\rho_{\rm oc}}}$. 

\begin{figure}
    \centering
    \includegraphics[width=0.65\linewidth]{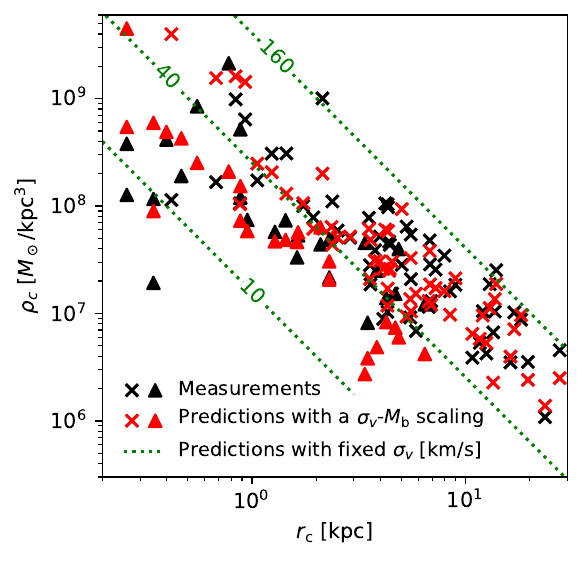}
    \caption{Comparison of the predicted core density–radius relation to the best-fit parameters given in \citet{Rodrigues:2017vto} (black crosses) and \citet{Oh:2015xoa} (black triangles). Red symbols are predictions with an empirically motivated scaling relation between the DM velocity dispersion $\sigma_v$ and the galactic baryon mass $M_{\rm b}$. Green dotted lines are predictions with some fixed DM velocity dispersion labeled in lines. }
    \label{fig:outer-core-density–radius}
\end{figure}

Numerically, we define $r_{\rm oc}$ to be the location where the logarithmic slope of the DM density profile is $-1.5$ to match the core radius in the Burkert profile \citep{Burkert:1995yz}, that is $\frac{d\ln\rho_{\rm d}}{d\ln r}\big\rvert_{r_{\rm{c}}}=-1.5$.  On the other hand, $\Delta_{\rm oc}$ is obtained by fitting Eq.~\eqref{eq:solution-outer-core} to the intermediate region near the outer core in the numerical solution of the DM profile. Subsequently, the outer core density $\rho_{\rm oc}$ is computed using Eq.~\eqref{eq:rho-SM} with the fitted $\Delta_{\rm oc}$. With these definitions and from a wide range of numerical solutions, we find that $\rho_{\rm oc}$ and $r_{\rm oc}$ are correlated by,
\begin{equation}\label{eq:rhooc-roc-relation}
    \rho_{\rm oc} = 1.4\times10^8\left(\frac{\sigma_v}{30\,{\text{km/s}}}\right)^2\left(\frac{1\,{\text{kpc}}}{r_{\rm oc}}\right)^2\,\frac{M_\odot}{{\rm kpc}^3}\,.
\end{equation}
This relation contains no other free parameter, such as the DM mass, as long as the outer core is induced. We compare the above prediction with the best-fit core densities and radii of a Burkert profile fitted to the rotation curves of LSB galaxies \citep{Rodrigues:2017vto} and dwarf galaxies \citep{Oh:2015xoa}.\footnote{Note that \citet{Oh:2015xoa} performed the analysis with a pseudo-isothermal profile instead of a Burkert profile. So, we have scaled the measured cored radii provided in \citet{Oh:2015xoa} by a factor of $\sqrt{3}$ to match the definition of the core radius here. We also have removed the data point of galaxy Haro 36 as the uncertainty of the core radius is above 100\%.} The comparison is shown in Figure~\ref{fig:outer-core-density–radius}. Remarkably, with $\sigma_v\sim40$\,km/s motivated by the stellar velocity dispersion of the dwarf galaxies with a similar baryon mass range, the predicted $\rho_{\rm oc}$-$r_{\rm oc}$ relation (green lines) matches the measurements (black symbols) very well. 

The predicted $\rho_{\rm oc}$-$r_{\rm oc}$ relation with a constant $\sigma_v$ is somewhat steeper than the measurements, i.e., the absolute slope is larger. This is because the variation of $\sigma_v$ with the galaxy size has not been taken into account. To do that, we assume the following scaling relation between $\sigma_v$ and $M_{\rm b}$ motivated by the empirical Faber-Jackson relation for dispersion-supported galaxies,
\begin{equation}\label{eq:sigma_v-Mb-scaling}
    \sigma_v=\left(\frac{M_{\rm b}}{10^9\,M_\odot}\right)^{1/4}\sigma_v^{\rm ref}\,,
\end{equation}
where $M_{\rm b}$ is the baryon mass of the galaxy and $\sigma_v^{\rm ref}$ is some reference DM velocity dispersion at $M_{\rm b}=10^9M_\odot$. We first take $\sigma_v^{\rm ref}=40$ km/s motivated by the stellar velocity dispersion for a dwarf galaxy of similar size \citep{binney2011galactic}. Note that the Faber-Jackson relation is between the luminosity (or the baryon amount) and the stellar velocity dispersion for dwarf galaxies. However, it is reasonable to assume that the invisible DM velocity dispersion is correlated with the baryon amount similarly.

For each galaxy, we use the galactic baryon mass (the sum of the masses of disk, bulge, and gas) given in \citet{Rodrigues:2014xka,Rodrigues:2017vto,Oh:2015xoa} to infer the DM velocity dispersion according to Eq.~\eqref{eq:sigma_v-Mb-scaling}. Then, we take the measured core radius and the inferred $\sigma_v$ as the inputs of Eq.~\eqref{eq:rhooc-roc-relation} to predict the core density. The results are shown by the red symbols in Figure~\ref{fig:outer-core-density–radius}. The predictions exhibit a good consistency with the measurements in both the slope and the magnitude. We quantify the goodness of fit using the coefficient of determination,
\begin{equation}\label{eq:R2}
R^2 = 1 - \frac{\sum_i (y_i - y_{{\rm fit},i})^2}{\sum_i (y_i - \bar{y})^2}\,,
\end{equation}
where $y_i$ denotes the measured core density (in logarithmic scale) for the ith galaxy, $y_{{\rm fit},i}$ is the predicted value from our model, and $\bar{y}$ is the mean of the measured core densities. We vary $\sigma_v^{\rm ref}$ find that the best fitting parameter is $\sigma_v^{\rm ref}=43$ km/s. To our knowledge, no alternative scenario has achieved such close agreement with the data in terms of the magnitude and slope in the core density–radius plane.

Our result also suggests that the observed $\rho_{\rm c}$-$r_{\rm c}$ relation should fundamentally be a $\rho_{\rm c}$-$r_{\rm c}$-$\sigma_v$ relation. Or, in terms of observables, it is a $\rho_{\rm c}$-$r_{\rm c}$-$M_{\rm b}$ relation assuming an empirically motivated scaling relation between the DM velocity dispersion and the galactic baryon mass. Only viewed in the projected $\rho_{\rm c}$-$r_{\rm c}$ plane, the product $\rho_{\rm c}r_{\rm c}$ is nearly (but not exactly) constant as reported in the literature \citep{2009MNRAS.397.1169D,Burkert:2015-scaling,Salucci:2018hqu}. Regarding this, by matching the outer core surface density with that reported in the observation ($\Sigma_0$), it was noted that $T_{\rm d}/m_{\rm d}=0.719 \sqrt{\Sigma_0 M_{\rm h}}$ \citep{Chavanis:2021jds}, where $M_{\rm h}$ is the DM mass within the outer core. This has a key difference from our Eq.~\eqref{eq:sigma_v-Mb-scaling}: our Eq.~\eqref{eq:sigma_v-Mb-scaling} is motivated by the well-established Faber-Jackson relation and involves the baryon mass, which is directly relates to the observable $M_{\rm b}$ instead of nonobservable $M_{\rm h}$.

We emphasize that the $\rho_c$–$r_c$ relation predicted by King-type cores has a physical origin distinct from that of the empirical Faber–Jackson relation. As shown above, even adopting a constant and reasonable value of $\sigma_v$ already reproduces the observed $\rho_c$–$r_c$ trend quite well, indicating that the King-type core captures the primary physical properties of DM cores. The Faber–Jackson relation is therefore not used to define or derive the $\rho_c$–$r_c$ relation, but enters our analysis only as a supporting empirical input to illustrate how variations in $\sigma_v$ across galaxies may modulate/improve the core properties predicted by King-type cores. In this sense, the role of the Faber–Jackson relation is supplementary rather than fundamental. The physical origin of the Faber–Jackson relation remains under active debate within the DM framework, but is likely connected to processes operating on larger scales than those directly addressed by the DID mechanism.

We also note that, the scaling relation Eq.~\eqref{eq:rhooc-roc-relation} applies to DM halos with resolved King-type cores. It does not necessarily means that all DM halos are cored. The diversity of DM inner profiles will be discussed in Sec.~\ref{sec:DID-hierarchical-formation}.

However, rather than the outer core itself, the most important feature of the FIP is the sharp density drop just outside the inner core, as we discuss below.\footnote{The gravitational effects of the region with a sharp density drop were examined previously focusing on the rotation curves \citep{RAR-2015}. In contrast, as we shall discuss, our work builds on a major reinterpretation: we highlight that the essential role of this feature is the density drop itself, which indirectly leads to the observable core especially in the paradigm of hierarchical cosmic structure formation. This stands in clear distinction from earlier interpretations.}

\section{The Degeneracy-Induced Depletion mechanism}\label{sec:the-DID-mechanism}
The agreement between the predictions and the measurements shown in Figure~\ref{fig:outer-core-density–radius} does not directly support the interpretation that the outer cores of fermionic dark matter halos correspond to the observed cores of dwarf galaxies and LSB galaxies. Instead, it suggests that the observed cores more closely resemble a King-type radius—the core scale of a classical self-gravitating isothermal sphere with an imposed finite central density \citep{S.W.cosmo.}. In fact, even the classical solution with $\Delta_0 < 0$ produces a core density–radius relation nearly identical to that given in Eq.~\eqref{eq:rhooc-roc-relation}. However, this classical cored isothermal sphere is known to suffer from gravitational instability \citep{1990PhR...188..285P}, rendering it physically unrealistic.

Nonetheless, the resemblance between the King radius and the observed DM halo cores offers a fresh angle on the cusp–core problem. Rather than requiring a mechanism that directly generates a cored DM profile, it is more plausible that some process first depletes DM in near the central region of galaxy and then the self-gravity of the low-density region subsequently configures it into a King-type core. This role is naturally fulfilled by the DID mechanism.

Unlike the classical cored isothermal halo, which imposes a finite central density by hand, the outer core of an FIP is induced by the presence of a degenerate inner core. Intuitively, for an FIP with a central degeneracy, the sharp drop in density just outside the inner core makes the right-hand side of Eq.~\eqref{eq:Hydro-equil} very small. This allows the pressure—and consequently the density—to vary slowly over a wide radius range. The sharp density drop works as follows: There is no centrally diverging solution to Eq.~\eqref{eq:Delta-differential}, indicating that the degenerate pressure is sufficiently stiff ($P_{\rm d}\propto \rho_{\rm d}^{5/3}$) to support a stable inner core against gravitational collapse. At the edge of the inner core, the degenerate DM rapidly transitions to a classical where $P_{\rm d}\propto\rho_{\rm d}$. This abrupt reduction in pressure stiffness necessitates a sharp drop in density to sustain a sufficient pressure gradient capable of counteracting gravity. Therefore, once a degenerate inner core forms, a steep drop in density at its boundary corresponding to the transition from degenerate to classical behavior is required.

A distinct and important feature of the DID mechanism is its counterintuitive response to gravitational collapse—opposite to the classical case—which sets it apart from other current proposals. In the classical case, the King radius contracts as the core density increases under gravitational collapse. In contrast, for FIPs, the density drop outside the inner core becomes more pronounced as the central density and degeneracy increase. This trend is evident in Figure~\ref{fig:mass-profiles}, where increasing central degeneracy $\Delta_0$ leads to more negative values of $\Delta_{\rm oc}$, corresponding to outer cores with lower densities. As more DM collapses into the degenerate inner core, the FIP becomes increasingly centrally dense with higher degeneracy, resulting in a more diluted and extended outer core.

The DID mechanism can be summarized as follows: \emph{The onset of degeneracy in fermionic dark matter induces a sharp drop in density beyond a compact degenerate inner core, preventing the dark matter from reaching high densities on larger scales.}

\section{DID mechanism incorporated with the hierarchical structure formation}\label{sec:DID-hierarchical-formation}
When the hierarchical cosmic structure formation paradigm is incorporated with the DID mechanism, the internal structure of DM halos deviates qualitatively from what described above and differs substantially from those predicted by classical N-body simulations. 

\begin{figure}
    \centering
    \includegraphics[width=0.65\linewidth]{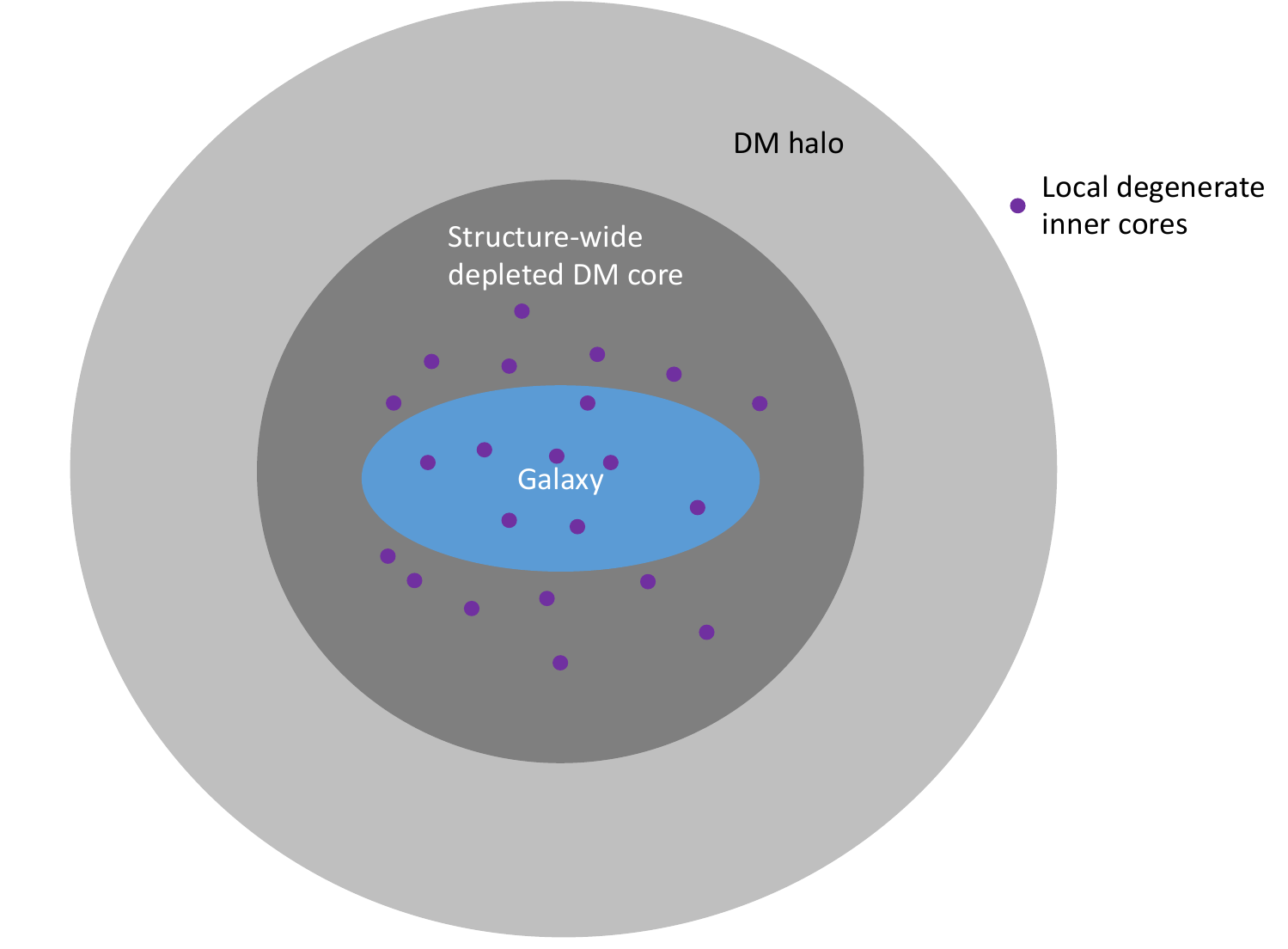}
    \caption{Schematic illustration of the proposed structure of a cored DM halo under the DID mechanism within the hierarchical cosmic structure formation paradigm. Compact degenerate inner cores (purple dots) form early within the smallest subhalos during hierarchical assembly. These local degeneracies deplete the surrounding DM, collectively producing a large-scale, low-density core (dark gray) embedded within the broader dark matter halo (light gray). The luminous galaxy is shown in blue.}
    \label{fig:schematic}
\end{figure}

Cosmic structure formation proceeds in a bottom-up manner: small-scale structures form first and gradually merge over cosmic time to build larger systems~\citep{1978MNRAS.183..341W,Springel-Frenk-White:2006}. A DM halo consists of numerous subhalos that collapsed earlier in the cosmic timeline, each of which can host a degenerate inner core. These local degeneracies deplete the surrounding DM and collectively prevent DM from forming galaxy-scale high-density regions. Although the inner cores are dense, their overall contribution to the halo mass is negligible. Recall that the self-gravity of a low-density region in equilibrium necessitates a King-type core. Once such a collective region forms, its self-gravity dominates and shapes the overall halo structure. The observable core of the host DM halo thus arises as a collective envelope formed by the overlapping outer regions of these early formed subhalos. Therefore, \textit{the key structural role of FIPs lies not in their outer cores per se, but in the density depletion induced by their degenerate inner cores.} Figure~\ref{fig:schematic} presents a schematic illustration of the structure of a cored DM halo under the DID mechanism incorporated in hierarchical cosmic structure formation paradigm. 
This proposed scenario opens a new avenue for addressing the observed diversity of DM inner profiles, which is currently difficult to reproduce consistently with baryonic feedback models \citep{Oman:2015xda,Relatores:2019-II,Sales:2022ich}. We discuss this point in detail below.

For a reasonable range of $\sigma_v$, the two key parameters governing FIP are the DM particle mass $m_{\rm d}$ and the central degeneracy parameter $\Delta_0$. When incorporated in the framework of hierarchical cosmic structure formation discussed above, however, it is not the $\Delta_0$ of an individual halo that is most relevant. Instead, the characteristic quantity is some average degeneracy of the constituent subhalos, denoted by $\langle \Delta_0 \rangle$.

The particle mass $m_{\rm d}$ sets the maximum density of the outer-core region of an individual FIP, which occurs in the case $\Delta_0 = 0$ (see Figure~\ref{fig:mass-profiles}), where the DID effect is absent. This density also represents the maximum attainable density of the King-type core collectively formed by the outer-core regions associated with local degeneracies. It therefore determines the minimum size of the King-type core through Eq.~\eqref{eq:rhooc-roc-relation} with the subscript ``oc" replaced with ``\textsc{kc}'' (King-type core), which is similar to Eq.~\eqref{eq:rs} for an individual FIP. When the DID mechanism is not efficient enough, i.e. when $\langle \Delta_0 \rangle$ is not sufficiently large, the resulting King-type core is dense and compact, effectively approaching a cuspy DM profile. Hence, for a given $m_{\rm d}$, the average degeneracy parameter $\langle \Delta_0 \rangle$ controls the size of the King-type core and governs whether a DM halo appears cuspy or cored.

To make the discussion quantitative, we first note that there exists a tight correlation between the central degeneracy parameter $\Delta_0$ and the outer–core degeneracy $\Delta_{\rm oc}$ for an individual FIP, which we determine numerically to be
\begin{equation}\label{eq:Delta0-Deltaoc}
 \Delta_{\rm oc} \simeq -0.743  \Delta_0 +0.033-\frac{20.2}{ \Delta_0}\,. 
\end{equation}
Thus, $\Delta_0$ and $\Delta_{\rm oc}$ are nearly linearly anticorrelated, especially in the regime of high central degeneracy. Again, within the hierarchical structure formation paradigm, $\Delta_0$ should be replaced by some average degeneracy of the constituent subhalos, denoted as $\langle\Delta_0\rangle$. Note that, at this stage, the averaging is meant in an effective sense, capturing the typical degeneracy of the constituent subhalos rather than a well-defined spatial or mass-weighted average. Combining Eq.~\eqref{eq:Delta0-Deltaoc} with Eq.~\eqref{eq:rho-SM}, we obtain King-type core density as
\begin{equation}\label{eq:outer-core-radius-Delta0}
\rho_{\textsc{kc}} \simeq 4.34\times10^{11}\,[M_\odot/{\rm kpc}^3] \times\left(\frac{m_{\rm d}}{1\,{\rm keV}}\right)^4\left(\frac{\sigma_v}{30\,{\rm km/s}}\right)^3\exp\left(-0.743 \langle \Delta_0 \rangle+0.033-\frac{20.2}{\langle \Delta_0 \rangle}\right)\,.
\end{equation}
The size of the King-type core can then be obtained by combining Eqs.~\eqref{eq:rhooc-roc-relation} and \eqref{eq:outer-core-radius-Delta0}, which increases with $\langle\Delta_0\rangle$ for a given $m_{\rm d}$.

For a DM particle mass consistent with cosmological constraints on warm dark matter (WDM), $m_{\rm d}\gtrsim5\,{\rm keV}$ [e.g., \citet{Irsic:2017yje,Dekker:2021scf}], the minimum size of the King-type core is $\sim0.23\,{\rm pc}$ according to Eq.~\eqref{eq:rs} when there is no DID effects, and becomes even smaller for larger $m_{\rm d}$. When DID becomes effective with an increasing positive $\langle\Delta_0\rangle$, the density of the King-type core decreases while its size increases. This implies that, for a given target King-type core density, there exists a required average degeneracy $\langle\Delta_0\rangle$ that approximately scales as
\begin{equation}
\langle\Delta_0\rangle_{\rm required}\propto5.4\ln m_{\rm d}
\end{equation}
when subdominant terms in Eq.~\eqref{eq:outer-core-radius-Delta0} are neglected and a reasonable value of $\sigma_v$ is considered. If $\langle\Delta_0\rangle<\langle\Delta_0\rangle_{\rm required}$, the King-type core is too compact to be observationally resolved, and the DM halo would appear cuspy. In contrast, when $\langle\Delta_0\rangle$ meets or exceeds the required value, the DM profile exhibits an observable core. Take $m_{\rm d}=5\,{\rm keV}$ for an example, achieving a King-type core density as low as $\rho_{\rm kc}=5\times10^8\,M_\odot/{\rm kpc}^3$—which roughly corresponds to the median density of observed DM cores with radii of $\sim1\,{\rm kpc}$ (Figure~\ref{fig:outer-core-density–radius})—requires $\langle\Delta_0\rangle_{\rm required}\simeq 17.4$ when adopting $\sigma_v=40\,{\rm km/s}$, with even larger values required for higher $m_{\rm d}$.

On the one hand, dwarf galaxies with similar stellar masses, $M_*\gtrsim10^7\,M_\odot$, can exhibit either cored or cuspy DM profiles; see \citet{Sales:2022ich} and references therein. Within the DID framework incorporated in hierarchical cosmic structure formation, it is the average degree of central degeneracy of the constituent subhalos, $\langle\Delta_0\rangle$, that determines whether a DM halo appears cuspy or cored. Because $\Delta_0$ is tied to the central density, or equivalently the degree of collapse, of the inner cores of the subhalos, it is naturally expected to correlate with the formation history of the host halo. Smaller host halos contain fewer subhalos and are therefore expected to exhibit a larger scatter in $\langle\Delta_0\rangle$, which in turn leads to a greater diversity of DM profiles.

On the other hand, there exists a clear correlation between galaxy size and core size: larger galaxies tend to host more extended DM cores. This trend also follows from the hierarchical formation scenario. Small subhalos that collapse at earlier epochs develop more larger inner degeneracies. Consequently, larger host halos, which assemble from a greater number of such early-forming subhalos, are intrinsically more likely to display extended and readily observable cores. However, it does not mean that all large DM halos—such as those in massive galaxies or clusters—necessarily exhibit cores large enough to be detectable. Within finite cosmic time, these subhalos may not have collapsed sufficiently to produce cores large enough for current observational sensitivity. Nonetheless, interestingly, the presence of even small-mass cores in massive halos can have significant implications—for example, in measuring the Hubble constant via time-delay strong lensing~\citep{Birrer:2020tax}.

\section{The stability of the DID mechanism against baryons' gravity}\label{sec:stability-against-baryon}
Thus far, we have considered DM-only halos. However, in realistic galactic environments, baryons are expected to condense toward the center and can significantly modify the gravitational potential in the inner regions. 
The inclusion of baryons modifies the Poisson equation, which is now expressed as:
\begin{equation}\label{eq:Poisson-baryon}
    \frac{d\phi}{dr}=\frac{G\left[M_{\rm d}(r)+M_{\rm b}(r)\right]}{r^2}\,.
\end{equation}
All other assumptions remain unchanged, i.e., Eqs.~\eqref{eq:Hydro-equil}–\eqref{eq:pressure-SM} remain the same. Note that we omit subscripts for the chemical potential $\mu$ and the variable $\Delta$, which refer specifically to DM, since only macroscopic quantities are considered for baryons. In particular, with the Thomas-Fermi approach that assumes $\mu$ is constant and using Eqs.\,\eqref{eq:Hydro-equil} and \eqref{eq:TF-approximation}, we again deduce that $T_{\rm d}$ is spatially independent. To model the baryonic mass distribution, we adopt the Hernquist profile  \citep{Hernquist:1990}:
\begin{equation}\label{eq:hernquist-profile}
    \rho_{\rm b}= \frac{M_{\rm b}^{\rm tot}}{2\pi r_{\rm s}^3}\frac{1}{\frac{r}{r_{\rm s}}(1+\frac{r}{r_{\rm s}})^3}\,,
\end{equation}
where $M_{\rm b}^{\rm tot}$ is the total baryon mass, and $r_{\rm s}$ represents the scale radius. Substituting Eq.~\eqref{eq:Poisson-baryon} into Eq.~\eqref{eq:Hydro-equil}, taking the derivative, and applying the baryon mass profile in Eq.~\eqref{eq:hernquist-profile}, we derive the following modified differential equation for $\Delta$, with again $r$ normalized by the characteristic radius $r_*$:
\begin{equation}\label{eq:Delta-differential-baryon} 
    \Delta''+\frac{2}{z}\Delta'=-J_{\nicefrac{3}{2}}^{\textsc{f}}(\Delta)+S_{\rm b}(z)\,.
\end{equation}
The source term $S_{\rm b}$, which accounts for the baryons, is given by
\begin{equation}\label{eq:source-baryon}
    S_{\rm b} = -\frac{\rho_{\rm b}}{\sqrt{2}Q_*\sigma_v^3} = -\frac{s_0}{\frac{z}{z_{\rm s}}(1+\frac{z}{z_{\rm s}})^3}\,,
\end{equation}
where $z_{\rm s}\equiv r_{\rm s}/r_*$ and $s_0 \equiv \frac{M_{\rm b}^{\rm tot}r_*^2\ell_{\rm p}}{4\pi m_{\rm p}r_{\rm s}^3\sigma_v^2}=\frac{\pi M_{\rm b}^{\rm tot}}{\sqrt{2}g\sigma_v^3r_{\rm s}^3m_{\rm d}^4}$ which reads
\begin{equation}\label{eq:s0}
\begin{split}
    s_0=5.14&\times10^{-7}\left(\frac{5\,\text{keV}}{m_{\rm d}}\right)^4\left(\frac{M_{\rm b}^{\rm tot}}{10^9M_\odot}\right) \\ &~~\times\left(\frac{1\,\text{kpc}}{r_s}\right)^3\left(\frac{30\,\text{km/s}}{\sigma_v}\right)^3\,,
\end{split}
\end{equation}
where we have used Eq.~\eqref{eq:rs} and taken $g=2$.

\begin{figure}
    \centering
    \includegraphics[width=0.65\linewidth]{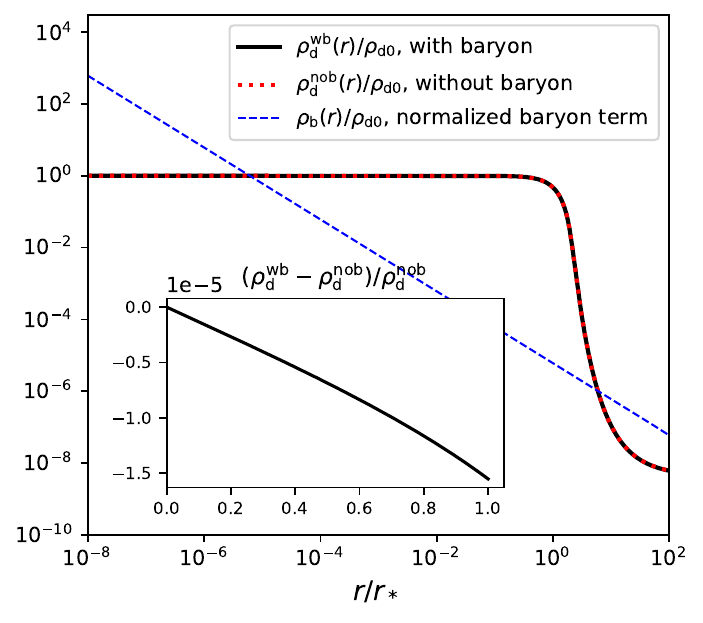}
    \caption{The effect of baryons in the inner core is negligible. The plot shows the normalized DM density profile for two cases: with baryons (solid curve) and without baryons (dotted curve). The normalized baryon source term is shown by the dashed curve. Since the overall density profiles near the inner core region are nearly indistinguishable, the inset plot highlights the fractional density difference between the two cases. This sample solution uses $m_{\rm d} = 5\,\mathrm{keV}$, $\Delta_0=20.85$, $r_{\rm s} = 1.8\,\mathrm{kpc}$, and $M_{\rm b}^{\rm tot} = 10^9\,M_\odot$. The velocity dispersion $\sigma_v$ is calculated using Eq.~\eqref{eq:sigma_v-Mb-scaling}.}
    \label{fig:b_effect_inner}
\end{figure}

There are two regions where $S_{\rm b}$ is larger than $J_{\nicefrac{3}{2}}^{\textsc{f}}(\Delta)$ and the presence of baryons might affect the solution of $\Delta$.

\subsection{Persistence of on the inner core against dense baryon environments}
First, when $z\rightarrow0$, $S_{\rm b}$ diverges as $S_{\rm b}\rightarrow -s_0z_{\rm s}/z$, which could dominate the right-hand side of Eq.~\eqref{eq:Delta-differential-baryon}. An important question is whether the inner core structure is modified by the diverging baryon density profile. This is essential as the persistence of the inner core is crucial for the DID mechanism.

Since $J_{\nicefrac{3}{2}}^{\rm F}\rightarrow\frac{2}{3}\Delta^{3/2}$ for $\Delta\gg1$ \citep{Zhang:2024vkx}, comparing the two terms on the right-hand side of Eq.~\eqref{eq:Delta-differential-baryon} it follows that $S_{\rm b}$ dominates in the region where $z<z_{\rm bd}$ with
\begin{equation}\label{eq:z_bd}
    z_{\rm bd} = \frac{3s_0z_s}{2\Delta_0^{3/2}}\,.
\end{equation}
The value $z_{\rm{bd}}$ is typically small given that $m_{\rm d}\gtrsim5$\,keV from cosmological constraints. For example, taking $r_{\rm oc}=1$\,kpc and $\sigma_v=30$\,km/s, Eq.~\eqref{eq:rs} gives $r_*<0.23$\,pc, while Eq.~\eqref{eq:outer-core-radius-Delta0} implies $\Delta_0>23$. Then from Eq.~\eqref{eq:z_bd}, $z_{\rm bd}<1.4\times10^{-5}$. The smallness of $z_{\rm bd}$ is due to the high density of the DM inner core, which dominates the compact inner region except at the very center.

For the baryon dominated region $z<z_{\rm bd}$, ignoring the $J_{\nicefrac{3}{2}}^{\textsc{f}}$ term, we can solve for $\Delta$, yielding
\begin{equation}\label{eq:Delta-small-r-baryon}
    \Delta = c_1 +\frac{c_2}{z}-\frac{s_0z_{\rm s}}{2}z\,.
\end{equation}
The term $\frac{c_2}{z}$ would destroy the inner core and cause the DM central density to diverge. However, this term contradicts the assumption that $J_{\nicefrac{3}{2}}^{\textsc{f}}$ can be neglected. Specifically, $J_{\nicefrac{3}{2}}^{\textsc{f}}\rightarrow\frac{2}{3}\Delta^{3/2}$ would scale as $\propto\frac{1}{z^{3/2}}$, diverging faster than the $S_{\rm b}$ term. Thus, the consistent solution includes only the constant term $c_1$ and the linearly decreasing term $-\frac{s_0z_{\rm s}}{2}z$. Since the region where Eq.~\eqref{eq:Delta-small-r-baryon} applies is small compared to the characteristic radius $r_*$, the decrease in $\Delta$ due to the $-\frac{s_0z_{\rm s}}{2}z$ term is negligible. Therefore, we can approximate $c_1\simeq\Delta_0$. In Figure~\ref{fig:b_effect_inner}, we show an example of numerical solutions with and without the baryon source starting with the same central value of $\Delta$. The difference between the two solutions for the inner core region is negligible. Physically, it means that the Fermi degenerate pressure can support the DM halo against the gravity generated by such a diverging baryon mass distribution. Note that the difference grows beyond $z_{\rm bd}$, which we shall explain in the next subsection.

The persistence of the FIP inner core highlights the robustness of the DID mechanism, since the existence of the inner core enforces a sharp density drop at its boundary as discussed earlier and shown in Figure~\ref{fig:b_effect_inner}.

\subsection{Impact on the outer core of FIP from baryons}\label{sec:baryon-impact-on-outer-core}
The second region where baryons may influence the system is the outer core. Importantly, any such effects do not compromise the DID mechanism, which relies solely on the persistence of the inner core. Gravitational effects from baryons would instead alter the detailed structure of the outer core region—more precisely speaking, the structure of the King-type core. Therefore, this serves more as a theoretical justification for the King-type core than for the DID mechanism itself.

Since we have shown that the effect in the inner region is negligible, we can first assume that the solution of $\Delta$ in the presence of baryons is nearly the same as the case without baryons and investigate under what conditions this assumption is broken. We denote the solution with baryons $\Delta_{\rm wb}$ and that without baryons $\Delta_{\rm nob}$, and we define their difference as $\delta\Delta\equiv\Delta_{\rm nob}-\Delta_{\rm wb}$. With assumption $\Delta_{\rm nob}\simeq\Delta_{\rm wb}$, we take the difference of Eq.~\eqref{eq:Hydro-equil} between the cases with and without baryons (and substituting Eqs.\,\eqref{eq:pressure-SM} and \eqref{eq:rho-SM}) gives 
\begin{equation}\label{eq:deltaDelta-diff}
    \frac{d(\delta\Delta)}{dr}= -\frac{GM_{\rm b}}{\sigma_v^2r^2}\,.
\end{equation}
The absence of the DM component means that the effect is insensitive to the DM properties, which is reasonable, since now the effect is mainly in the classical region. We can estimate the $|\delta\Delta|$ by
\begin{equation}\label{eq:deltaDelta_max}
    |\delta\Delta|\sim\int_0^\infty\frac{GM_{\rm b}}{\sigma_v^2r^2}dr\,.
\end{equation}
With $\rho_{\rm b}$ given by Eq.~\eqref{eq:hernquist-profile}, we have $M_{\rm b}=M_{\rm b}^{\rm tot}\frac{(r/r_{\rm s})^2}{(1+r/r_{\rm s})^2}$. Substituting it into Eq.~\eqref{eq:deltaDelta_max}, we obtain
\begin{equation}\label{eq:deltaDelta_max_hernquist}
    |\delta\Delta|\sim\frac{GM_{\rm b}^{\rm tot}}{\sigma_v^2 r_{\rm s}}\equiv \kappa_{\rm b}\,,
\end{equation}
where we call $\kappa_{\rm b}$ the baryon compactness.

\begin{figure}
    \centering
    \includegraphics[width=0.65\linewidth]{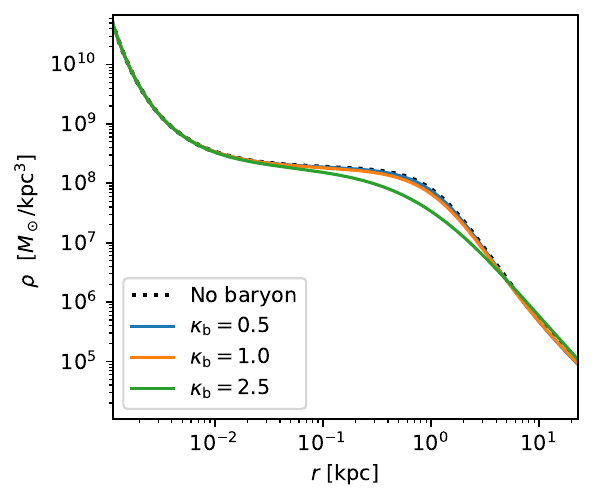}
    \caption{Effects of baryons on the DM outer core. Note that any such effect only alters the details of the outer core without jeopardizing the DID mechanism, as both the inner core and the sharp density drop remain stable. Significant alteration of the DM outer core occurs only when the parameter $\kappa_{\rm b} \gtrsim 1$ but $\kappa_{\rm b}<1$ is expected for dwarf-size galaxies.}
    \label{fig:b_effect_outer}
\end{figure}

Since now the change in the solution for $\Delta$ occurs in the classical regime where $\Delta \ll -1$, the DM density scales as $\rho_{\rm d} \propto \exp(\Delta)$. Consequently, a substantial modification of the DM density profile requires $\delta\Delta \sim \kappa_{\rm b}$ to be at least of $\mathcal{O}(1)$ or larger. This effect is illustrated in Figure~\ref{fig:b_effect_outer}, where the outer core profile exhibits significant alterations only for $\kappa_{\rm b} \gtrsim 1$.

We note that $r_{\rm s}$ is not well-defined due to the complexity of galaxies in question—each containing gas, disk, and bulge components—$\kappa_{\rm b}$, however, is generally expected to be less than unity for the following reason. If the system is baryon-dominated, the virial theorem suggests that $\frac{G M_{\rm b}^{\rm tot}}{\sigma_v^2 r_{\rm s}} \sim \mathcal{O}(1)$. When the gravitational potential is dominated by DM, the velocity dispersion $\sigma_v$ increases due to the deeper potential well, thereby reducing $\kappa_{\rm b}$. As a result, the outer core is not expected to be significantly affected by the baryonic gravitational potential. This also implies that in baryon-dominated systems—where the DID mechanism still operate—the resulting dark matter profiles could be more complex. A detailed investigation of such cases is beyond the scope of this work.

Besides the effects from the baryon potential, we remark on the potential impact of the central black hole (BH). If the BH is co-centered with the FIP, \ref{sec:impact-BH} shows that the inner core would be quickly destroyed through BH accretion. However, unlike baryonic matter—which extends across spatial scales comparable to that of DM—the influence of a central BH is confined to a very small region. Within the framework of hierarchical structure formation, to be explored in the next section, it is unlikely that local degeneracies would consistently align with the galactic center. Instead, the BH is likely embedded within the induced low-density DM core. As a result, the impact of the central BH, as in the case of conventional cored DM profiles assumed, is expected to be negligible.

\subsection{Rotation Curves and their Diversity}
\begin{figure}
    \centering
    \includegraphics[width=0.65\linewidth]{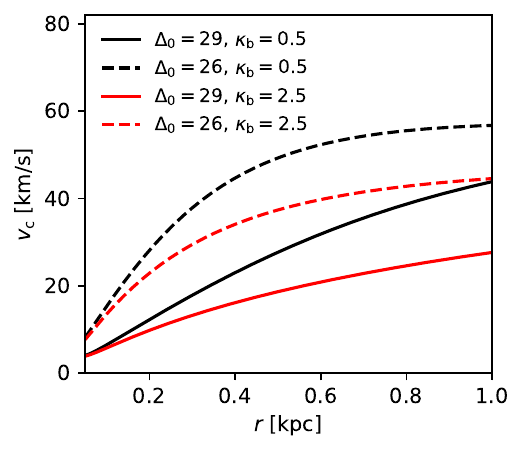}
    \caption{Diversity of rotation curves predicted from FIPs with different central degeneracy $\Delta_0$ and baryon compactness $\kappa_{\rm b}$. In the physical scenario proposed in this work, where the DID mechanism is incorporated into hierarchical cosmic structure formation, $\Delta_0$ should be interpreted as representing the average degeneracy $\langle\Delta_0\rangle$ of the constituent subhalos within a host DM halo. Halos with sufficiently large $\langle\Delta_0\rangle$ generate cored-halo rotation curves (solid lines), whereas those with insufficient $\langle\Delta_0\rangle$ produce cuspy-halo–like rotation curves (dashed lines). For physically reasonable values of $\kappa_{\rm b}$, of order unity or smaller, increasing $\kappa_{\rm b}$ tends to make the rotation curves shallower without significantly altering their overall characteristics. Here, we take $m_{\rm d}=30$ keV and $M_{\rm b}^{\rm tot}=5\times10^8\,M_{\odot}$.}
    \label{fig:rotation_curves}
\end{figure}

To better account for direct observables, we compute the rotation curves predicted from FIPs with different central degeneracy parameters $\Delta_0$ and different baryon compactness $\kappa_{\rm b}$, and present the results in Figure~\ref{fig:rotation_curves}. Here, $\Delta_0$ should be interpreted as representing the average degeneracy of the constituent subhalos $\langle\Delta_0\rangle$ within the hierarchical structure formation paradigm. Consequently, we show only the rotation curves associated with the King-type core region and larger radii, while excluding the inner degenerate core of an individual FIP, which does not correspond to the physical scenario considered in this work. For the results presented in Figure~\ref{fig:rotation_curves}, we take $m_{\rm d}=30$ keV and $M_{\rm b}^{\rm tot}=5\times10^8\,M_{\odot}$. A sufficient $\langle\Delta_0\rangle$ is chosen to be $29$ so that the King-type core size is approximately $1$ kpc. 

As discussed in Sec.~\ref{sec:DID-hierarchical-formation}, the average degeneracy $\langle\Delta_0\rangle$ controls the size of the King-type core, allowing a DM halo to appear cored when $\langle\Delta_0\rangle$ is sufficiently large, or cuspy when $\langle\Delta_0\rangle$ is insufficient and the resulting King-type core is too compact to be resolved. This diversity is illustrated in Figure~\ref{fig:rotation_curves} by the two black curves. For $m_{\rm d}=30\,{\rm keV}$, a value of $\langle\Delta_0\rangle\simeq29$ is required to generate a kpc-scale King-type core, in which case the corresponding rotation curve (solid black) is near linearly for $r\lesssim1\,{\rm kpc}$, as for the case of a cored DM halo. In contrast, when $\langle\Delta_0\rangle$ is insufficient (e.g., $\langle\Delta_0\rangle\simeq26$), the King-type core is too small, and the resulting rotation curve (dashed black) resembles that of a cuspy DM halo.

Different values of $\kappa_{\rm b}$ also contribute, to some extent, to the diversity of the rotation curves. As discussed in the previous subsection, physically meaningful values of $\kappa_{\rm b}$ are expected to be of order unity or smaller. In Figure~\ref{fig:rotation_curves}, we compare cases with $\kappa_{\rm b}=0.5$ where the King-type core is nearly unaffected, to cases with $\kappa_{\rm b}=2.5$ where the King-type core is moderately modified. Increasing $\kappa_{\rm b}$ tends to make the rotation curves shallower, while leaving their overall characteristics (in terms of linearity) largely unchanged.

\section{Conclusion}

In this work, we propose a Degeneracy-Induced Depletion mechanism that generates indirectly cored dark-matter halos and accounts for their diversity, offering a new perspective on fermionic isothermal halos. In this scenario, a sharp drop in the DM density is triggered by the presence of a degenerate inner core and remains stable even in dense baryonic environments. The role of this abrupt density decrease in shaping the profiles of DM halos has not been recognized in the existing literature. We propose that, within the bottom-up paradigm of cosmic structure formation, the smallest and earliest-collapsed constituent subhalos develop sufficient degeneracies. These degenerate subhalos collectively generate a galaxy-scale low-density region, which subsequently configure through self-gravity into a classical King-type core of the host DM halo. The resulting King-type cores, once present, exhibit a density–radius relation consistent with observations of dwarf and LSB galaxies. The average degeneracy of the constituent subhalos determines the size of the King-type core, allowing a DM halo to have a observable core or to appear cuspy when the King-type core is too compact. This scenario provides a new framework for understanding the observed diversity of DM inner profiles in dwarf-sized galaxies, potentially linking it to their formation histories.

This scenario predicts the presence of compact, degenerate constituent cores within a host dark-matter halo, corresponding to the smallest subhalos that collapsed early in cosmic history and thus developed sufficient degeneracies. These local degenerate subhalos behave as MACHO-like cores (Massive Compact Halo Objects). Assuming that the number of such degenerate cores is comparable to the abundance of the smallest subhalos found in conventional DM-only $N$-body simulations, a Milky Way–sized galaxy could host thousands of those MACHO-like cores. Given the WDM constraint $m_{\rm d}\gtrsim5\,{\rm keV}$, the masses of these degenerate cores can reach up to $\sim2\times10^6\,M_\odot$ [see Eq.~\eqref{eq:inner-core-mass} in the Appendix], with the lower mass limit primarily set by the upper bound on the DM particle mass. This prediction offers a compelling opportunity to test the DID scenario through lensing observations in ongoing and upcoming wide-field surveys, including the CSST (Chinese Space Station Telescope), the Vera C. Rubin Observatory, the Nancy Grace Roman Space Telescope, and Euclid.

A caveat is that the relationship between local degeneracies, the size of the induced DM core, and the formation history of DM halos remains unclear. This connection is fundamentally tied to the dark matter particle mass: if the mass is too large, the required local degeneracies may not be achievable within the finite age of the Universe. A more complete understanding of this interplay would help refine predictions for the size and abundance of MACHO-like inner cores. A detailed investigation of these questions is left to future work.

Our results suggest that cored dark matter halos may naturally arise from fermionic quantum statistics, indirectly but robustly through the DID mechanism, without invoking exotic self-interactions or violent baryonic feedback. The DID mechanism offers a new theoretical pathway toward understanding the formation of low-density cores and their dependence on halo assembly history. These findings imply that a complete understanding of cosmic structure on small scales will require future simulations to incorporate fermionic quantum statistics. While current N-body simulations neglect degeneracy pressure, our results provide a physically motivated framework for modeling such effects analytically and for guiding the development of new simulation techniques—potentially via effective pressure terms or phase-space-limited dynamics—to capture quantum statistical behavior. Our results also open a new window for probing the fundamental properties of DM through astrophysical observations.

\begin{acknowledgements}
We thank the anonymous referee for carefully reviewing of our manuscript and for providing helpful comments and suggestions. W. L. thanks P. H. Chavanis, Y. Cui and T. T. Yanagida for the useful feedback. W. L. acknowledges that this work is supported by the ``Science \& Technology Champion Project" (202005AB160002), the ``Top Team Project" (202305AT350002) and the ``Innovation Team Project'' (202105AE160021), all funded by the ``Yunnan Revitalization Talent Support Program." This work is also supported by the ``Yunnan Key Laboratory of Survey Science" (202449CE340002), the ``Yunnan General Grant'' (202401AT070489) and the National Key R\&D Program of China (2024YFA1611600).
\end{acknowledgements}

\appendix
\section{Derivation of the differential equation for the degeneracy parameter}\label{sec:derivation}
The second-order differential equation Eq.~\eqref{eq:Delta-differential} fully describes FIP. To derive this equation, we first show that the Thomas--Fermi approach with a spatially independent real chemical potential is equivalent to the isothermal condition. With $\mu$ being constant, we have
\begin{equation}\label{eq:phi-derivative}
\frac{d\phi}{dr} = -\frac{d}{dr}\left(\frac{\mu_{\rm eff}}{m_{\rm d}}\right)
= -\frac{d}{dr}\left(\frac{T_{\rm d}}{m_{\rm d}} \Delta\right)\,,
\end{equation}
Substituting this together with Eqs.~\eqref{eq:rho-SM} and \eqref{eq:pressure-SM} into Eq.~\eqref{eq:Hydro-equil}, we obtain
\begin{equation}\label{eq:Tprime}
\Big[ 5J^{\rm F}_{5/2}(\Delta) - 3\Delta J^{\rm F}_{3/2}(\Delta) \Big]
\frac{d}{dr}\left(\frac{T_{\rm d}}{m_{\rm d}}\right) = 0\,,
\end{equation}
where we have used $\frac{d}{d\Delta}J^{\rm F}_{5/2}(\Delta)=\frac{3}{2}J^{\rm F}_{3/2}(\Delta)$. Since the prefactor in brackets is always positive, it follows that
\begin{equation}
\frac{d}{dr}\left(\frac{T_{\rm d}}{m_{\rm d}}\right) = 0 \, ,
\end{equation}
implying that the dark matter temperature is also spatially constant.

With the DM temperature being constant, we substitute the Poisson equation, Eq.~\eqref{eq:Poisson-eq}, into Eq.~\eqref{eq:Hydro-equil}, multiply by $r^2$, and then take the derivative with respect to $r$. This gives
\begin{equation}\label{eq:intermediate-step}
    \frac{d}{dr}\!\left( \frac{r^2}{\rho_{\rm d}} \frac{dP_{\rm d}}{dr} \right)
    = - 4\pi G \, \rho_{\rm d} r^2 \, .
\end{equation}
Finally, by substituting Eqs.~\eqref{eq:rho-SM} and \eqref{eq:pressure-SM} into the above and defining the dimensionless variable 
\[
z \equiv \frac{r}{r_*}, \qquad r_* \ \text{given by Eq.~\eqref{eq:rs}} \, ,
\]
we obtain the second-order differential equation
\begin{equation}\label{eq:FIP-ODE}
    \frac{d^2 \Delta}{dz^2} + \frac{2}{z} \frac{d\Delta}{dz}
    = - J^{\rm F}_{3/2}(\Delta) \, .
\end{equation}

\section{Impact of the central black hole on co-centered FIPs}\label{sec:impact-BH}
Here, we investigate the effects of a co-centered black hole (BH) on FIPs. 

Naively, attempting to find an equilibrium configuration of an FIP with a co-centered BH leads to a divergent DM density at the center. This divergence arises when the BH gravitational potential is included in the Poisson equation [Eq.~\eqref{eq:Poisson-eq}]; integrating the hydrostatic equilibrium equation [Eq.~\eqref{eq:Hydro-equil}] while neglecting the DM self-gravity term yields a DM density that diverges as $r \rightarrow 0$. However, such intense gravitational influence at small scales suggests that the equilibrium assumption breaks down in this regime. A more appropriate question is whether BH accretion significantly modifies the DM halo structure.

We model this effect using Bondi accretion, embedding the BH within the inner core of the FIP. Recall that the Bondi accretion rate for a BH with mass $M_{\rm BH}$, embedded in an otherwise uniform-density medium with density $\rho_\infty$ and adiabatic index $\gamma$, is given by  \citet{1952MNRAS.112..195B}:
\begin{equation}\label{eq:bondi-accretion}
\dot{M}_{\rm BH} = \lambda(\gamma)\frac{4\pi G^2 M_{\rm BH}^2 \rho_\infty}{c_{\rm s,\infty}^3},\,
\end{equation}
where $c_{\rm s,\infty}$ is the sound speed of the medium without the BH, and $\lambda(\gamma)$ is a dimensionless factor that depends on the adiabatic index of the accreting gas. Here, $\rho_\infty= \rho_0$ and, since the specific entropy is only a function of $\Delta$ \citep{Lin:2023fao}, from Eqs.~\eqref{eq:rho-SM} and \eqref{eq:pressure-SM} the adiabatic (isentropic) sound speed is always, 
\begin{equation}\label{eq:addiabatic-sound-speed}
    c_{\rm s} = \sqrt{\frac{5}{3} \frac{P_{\rm d}}{\rho_{\rm d}}}\,,
\end{equation}
which is $\sqrt{\frac{2\Delta_0}{3}}\sigma_v$ in the inner-core region. The adiabatic index is thus $5/3$ independent of $\Delta$, which gives $\lambda(\gamma=5/3) =0.25$ \citep{1952MNRAS.112..195B}. At the same time, the inner core is embedded in and accreting DM from the outer core at a rate given by
\begin{equation}\label{eq:inner-core-accretion-rate}
    \dot{M}_{\rm ic} = \lambda(\gamma_{\rm oc}) \frac{4\pi G^2 M_{\rm ic}^2 \rho_{\rm oc}}{c_{\rm s,oc}^3}\,.
\end{equation}
The sound speed in the outer core is
 $c_{\rm s,oc} = \sqrt{\frac{5}{3} \frac{P_{\rm oc}}{\rho_{\rm oc}}} = \sqrt{\frac{5}{3}}\, \sigma_v$. 

In order for the inner core to survive, the net mass change since the galaxy’s formation (taken to be at redshift $z = 7$, corresponding to a cosmic lookback time of $t_{\rm tot} \simeq 10~\mathrm{Gyr}$) must be less than the core’s initial mass $ M_{\rm ic}$. This gives the condition:
\begin{equation}\label{eq:BH-accretion-requirement1}
    (\dot{M}_{\rm BH} - \dot{M}_{\rm ic})\, t_{\rm tot} < M_{\rm ic}\,.
\end{equation}
Solve Eq.~\eqref{eq:Delta-differential} near the inner core region, it can be shown that for a sufficiently high central degeneracy $\Delta_0\gg1$, the inner core density profile is approximately 
\begin{equation}\label{eq:inner-core-density}
    \rho_{\rm ic}\simeq\rho_0\exp(-r^2/r_{\rm ic}^2)\,,
\end{equation}
where the inner core radius is given by,
\begin{equation}\label{eq:inner-core-radius}
    r_{\rm ic}=\frac{\sqrt{6}}{\Delta_0^{1/4}}r_*\,.
\end{equation}
Then, the mass of the inner core is 
\begin{equation}\label{eq:inner-core-mass}
    M_{\rm ic}=4\pi\rho_{\rm ic} r^2dr\simeq(6\pi)^{3/2}\Delta_0^{-3/4}r_*^3\rho_0\,.
\end{equation}
Combining this with Eq.~\eqref{eq:BH-accretion-requirement1}, we obtain:
\begin{equation}\label{eq:mass-accretion-BH-max}
M_{\rm BH}<\sqrt{M_1^2+M_2^2}\,,
\end{equation}
where 
\begin{align}
    M_1=&\left(\frac{8\pi^{1/2}\Delta_0^{3/4} r_*^3 \sigma_v^3}{G^2 t_{\rm tot}}\right)^{1/2} \\
    M_2=&M_{\rm ic}\left(\frac{J_{\nicefrac{3}{2}}^{\textsc{f}}(\Delta_{\rm oc})}{J_{\nicefrac{3}{2}}^{\textsc{f}}(\Delta_0)}\right)^{1/2}\left(\frac{2\Delta_0}{5}\right)^{3/4}
\end{align}

As previously noted, a fermionic DM candidate that satisfies the WDM cosmological constraint $m_{\rm d} \gtrsim 5\,\mathrm{keV}$ implies a central degeneracy parameter of $\Delta_0 \gtrsim 19$.
Taking $m_{\rm d} = 5\,\mathrm{keV}$, $\Delta_0 \gtrsim 19$, and a velocity dispersion of $\sigma_v = 40\,\mathrm{km/s}$, we compute the DM density profile and extract the quantities needed to evaluate the enclosed masses $M_1$ and $M_2$. This analysis yields an upper bound on the central black hole mass:
\begin{equation}\label{eq:BH-mass-limit1}
    M_{\rm BH} \lesssim 800\,M_{\odot}\,.
\end{equation}
This limit becomes increasingly stringent for larger values of the dark matter particle mass $m_{\rm d}$, further constraining the allowed BH mass. 

The above upper bound is even lower than that of some confirmed BHs in the galaxies under consideration, which would falsify FIPs as a solution to the cusp–core problem. However, as noted earlier, when the DID mechanism is incorporated into the hierarchical structure formation paradigm, BHs are unlikely to be co-centered with the local degenerate inner cores. Instead, they are typically embedded in the more extended, low-density outer core regions where the accretion rate is much lower. In this scenario, the effect of the central BH on the induced cored DM halo would be equivalent to that in the conventional case where an empirical cored dark matter profile is assumed (e.g., the Burkert profile). Consequently, the mass limit given in Eq.~\eqref{eq:BH-mass-limit1} becomes essentially irrelevant.

\section{Comments on general relativistic correction}
We have assumed that general relativistic (GR) corrections to FIPs are small, as previously shown in \citet{RAR-2015}. Here, we provide an explicit demonstration from a different perspective. For GR effects to become significant, the compactness of an FIP must be sufficiently large. The maximum compactness occurs at the edge of the inner core, which, using Eqs.~\eqref{eq:inner-core-radius} and \eqref{eq:inner-core-mass}, is given by
\begin{equation}\label{eq:largest-FIP-compactness}
\frac{M_{\rm ic}}{r_{\rm ic}} = \sqrt{\pi}\Delta_0 \sigma_v^2\,.
\end{equation}
For GR corrections to be negligible, we require $\Delta_0 \sigma_v^2 \ll 1$. This same condition coincides with the regime in which dark matter is degenerate but remains nonrelativistic \citep{Zhang:2024vkx}, as it should be. In practice, typical values are $\sigma_v \sim 10^{-4}$ and $\Delta_0 < 100$. Achieving substantially larger $\Delta_0$ is unlikely, as we will demonstrate in an upcoming work. Therefore, the condition $\frac{M_{\rm ic}}{r_{\rm ic}} \ll 1$ is easily satisfied, justifying our neglect of GR corrections.


\bibliography{references}
\bibliographystyle{aasjournalv7}

\end{document}